\documentstyle[epsf]{mn}

\newif\ifAMStwofonts

\ifoldfss
  \ifCUPmtlplainloaded \else
    \NewTextAlphabet{textbfit} {cmbxti10} {}
    \NewTextAlphabet{textbfss} {cmssbx10} {}
    \NewMathAlphabet{mathbfit} {cmbxti10} {} 
    \NewMathAlphabet{mathbfss} {cmssbx10} {} 
  \fi
  \ifAMStwofonts
    \ifCUPmtlplainloaded \else
      \NewSymbolFont{upmath} {eurm10}
      \NewSymbolFont{AMSa} {msam10}
      \NewMathSymbol{\upi}     {0}{upmath}{19}
      \NewMathSymbol{\umu}     {0}{upmath}{16}
      \NewMathSymbol{\upartial}{0}{upmath}{40}
      \NewMathSymbol{\leqslant}{3}{AMSa}{36}
      \NewMathSymbol{\geqslant}{3}{AMSa}{3E}

       \let\le=\leqslant
       
    \fi
  \fi
\fi 

\ifnfssone
  \newmathalphabet{\mathit}
  \addtoversion{normal}{\mathit}{cmr}{m}{it}
  \addtoversion{bold}{\mathit}{cmr}{bx}{it}
  \newmathalphabet{\mathbfit} 
  \addtoversion{normal}{\mathbfit}{cmr}{bx}{it}
  \addtoversion{bold}{\mathbfit}{cmr}{bx}{it}
  \newmathalphabet{\mathbfss} 
  \addtoversion{normal}{\mathbfss}{cmss}{bx}{n}
  \addtoversion{bold}{\mathbfss}{cmss}{bx}{n}
  \ifAMStwofonts
    \ifCUPmtlplainloaded \else
      \UseAMStwoboldmath
      \makeatletter
      \new@mathgroup\upmath@group
      \define@mathgroup\mv@normal\upmath@group{eur}{m}{n}
      \define@mathgroup\mv@bold\upmath@group{eur}{b}{n}
      \edef\UPM{\hexnumber\upmath@group}
      \new@mathgroup\amsa@group
      \define@mathgroup\mv@normal\amsa@group{msa}{m}{n}
      \define@mathgroup\mv@bold\amsa@group{msa}{m}{n}
      \edef\AMSa{\hexnumber\amsa@group}
      \makeatother
      \mathchardef\upi="0\UPM19
      \mathchardef\umu="0\UPM16
      \mathchardef\upartial="0\UPM40
      \mathchardef\leqslant="3\AMSa36
      \mathchardef\geqslant="3\AMSa3E

       \let\le=\leqslant

    \fi
  \fi
\fi 

\ifnfsstwo
  \DeclareMathAlphabet{\mathbfit}{OT1}{cmr}{bx}{it}
  \SetMathAlphabet\mathbfit{bold}{OT1}{cmr}{bx}{it}
  \DeclareMathAlphabet{\mathbfss}{OT1}{cmss}{bx}{n}
  \SetMathAlphabet\mathbfss{bold}{OT1}{cmss}{bx}{n}
  \ifAMStwofonts
    \ifCUPmtlplainloaded \else
      \DeclareSymbolFont{UPM}{U}{eur}{m}{n}
      \SetSymbolFont{UPM}{bold}{U}{eur}{b}{n}
      \DeclareSymbolFont{AMSa}{U}{msa}{m}{n}
      \DeclareMathSymbol{\upi}{0}{UPM}{"19}
      \DeclareMathSymbol{\umu}{0}{UPM}{"16}
      \DeclareMathSymbol{\upartial}{0}{UPM}{"40}
      \DeclareMathSymbol{\leqslant}{3}{AMSa}{"36}
      \DeclareMathSymbol{\geqslant}{3}{AMSa}{"3E}

       \let\le=\leqslant

    \fi
  \fi
\fi 

\ifCUPmtlplainloaded \else
  \ifAMStwofonts \else 
    \def\upi{\pi}
    \def\umu{\mu}
    \def\upartial{\partial}
  \fi
\fi

\title[KPD 0422+5421]{KPD 0422+5421: 
A New Short Period Subdwarf B/White Dwarf Binary}
\author[Koen et al.]{C. Koen$^{1,2}$, Jerome 
         A. Orosz$^{3}$\thanks{Guest 
Observer, McDonald Observatory, University of Texas at Austin}
\& Richard A. Wade$^{3}$\\
   $^1$South African Astronomical Observatory, P.O. Box 9, 7935 Observatory,
South Africa\\
     $^2$Department of Astronomy, University of Texas, Austin, TX 78712, USA\\
        $^3$Department of Astronomy \& Astrophysics, The Pennsylvania State
University, 525 Davey Laboratory, University Park, PA 16802, USA }
\date{\today}
\pagerange{\pageref{firstpage}--\pageref{lastpage}}
\pubyear{1998}

\begin{document}

\maketitle

\label{firstpage}

\begin{abstract}
The sdB star KPD 0422+5421 was discovered to be a single-lined
spectroscopic binary with a period of $P=0.0901795\pm (3\times
10^{-7})$ days (2 hours, 10 minutes).  The $U$ and $B$ light curves
display an ellipsoidal modulation with amplitudes of $\approx 0.02$
magnitudes.  The sdB star contributes nearly all of the observed flux.
This and the absence of any reflection effect suggest that the unseen
companion star is small (i.e.\ $R_{\rm comp}\approx 0.01\,R_{\odot}$)
and therefore degenerate.  We modeled the $U$ and $B$ light curves and
derived $i=78.05\pm 0.50^{\circ}$ and a mass ratio of $q=M_{\rm
comp}/M_{\rm sdB}=0.87\pm 0.15$.  The sdB star fills 69\% of its Roche
lobe.  These quantities may be combined with the mass function of the
companion ($f(M)=0.126 \pm 0.028\,M_{\odot}$) to derive $M_{\rm
sdB}=0.72\pm 0.26\,M_{\odot}$ and $M_{\rm comp}= 0.62\pm
0.18\,M_{\odot}$.  
We used model spectra to derive the effective temperature, surface gravity,
and helium abundance of the sdB star.  We found
$T_{\rm eff}=25,000\pm 1500$~K, $\log g=5.4\pm 0.1$, and [He/H] =
$-1.0$.  
With a period of 2 hours and 10 minutes, KPD
0422+5421 has one of the shortest known orbital periods of a detached
binary.  This system is also one of only a few known binaries which
contain a subdwarf B star and a white dwarf.  Thus KPD 0422+5421
represents a relatively unobserved, and short-lived, stage of binary
star evolution.
\end{abstract}


\section{INTRODUCTION}\label{intro}
The star KPD 0422+5421 was discovered by Downes (1986) during the
course of a search for very blue stars in the galactic plane, and
classified as an sdB star, i.e. a hot, hydrogen-rich subdwarf.
R. Saffer (private communication) derived from spectroscopic
measurements an effective temperature $T_{\rm eff}=26,050$ and a
surface gravity $\log g=5.51$; Saffer did not communicate a helium
abundance.  The sdB classification prompted one of the current authors
to include KPD 0422+5421 in a list of candidates to be monitored for
rapid pulsations (see e.g.\ Kilkenny 1997a). During the course of the
mandatory 90 minute high speed photometric run on the star, it
appeared to vary with a period of about an hour, at an amplitude of
roughly 0.01 mag.  Further spectroscopic and photometric observations
show that KPD 0422+5421 is a close binary star with a period of
$2.1643$ hours, one of the shortest known orbital periods among the
detached binaries.  We argue that the companion star is a white dwarf,
making KPD 0422+5421 one of a small number of known systems with a sdB
and a white dwarf.  We report here our observations, data reductions,
and data analysis.

\section{THE HIGH SPEED PHOTOMETRIC OBSERVATIONS AND ANALYSIS}\label{hsp}
All the photometry reported here was obtained at the University of
Texas' McDonald Observatory on Mt Locke, Texas.  Two different
instrumental configurations were used: the Louisiana State University
Photometer (P-LSU) mounted on the 0.9m telescope, and the Stiening
Photometer attached to the 2.1m telescope. The P-LSU is a standard
two-channel photoelectric photometer (see e.g. Grauer \& Bond 1981),
equipped with a blue-sensitive Hamamatsu R647 photomultiplier tube.
The Stiening Photometer is a four-channel instrument which allows
simultaneous high speed photometry in four different wavebands (Horne
\& Stiening 1985; Wood, Zhang \& Robinson 1993). The current passbands
of the instrument are similar to Johnson's $UBVR$, although there are
notable differences (Wood et al. 1993). One channel of a second
photometer (P45) was used as a fifth channel with the Stiening
photometer, to monitor a comparison star.  Observations with both
P-LSU and P45 photometers were in ``white light'', i.e.  no filters
were placed in the light beam. The resulting effective wavelengths are
similar to Johnson's $B$, but with substantially wider
bandpasses. Integration of 10 and 5 seconds were used when running
respectively the P-LSU and Stiening photometers.

\begin{table}
\centering
\begin{center}
\caption[]{Log of the observations, all of which were made at the
University of Texas' McDonald Observatory.  All 0.9m observations were
obtained under good photometric conditions, while all observations on
the 2.1m telescope were acquired under poor conditions. The
instrumental configurations are described in the text. JD
2,450,752.8437 is 1997 October 31.3438 UT}
\begin{tabular}{ccc} \hline
Starting Time &    Run Length & Telescope \\
JD 2,450,000+ &       (Hours)    &  (meters)  \\
\hline
752.8430 & 1.4  & 0.9 \\
753.8599 & 3.0  & 0.9 \\
754.7085 & 6.9 & 0.9  \\
785.7645 & 4.0 & 2.1 \\
787.7779 & 3.9 & 2.1 \\ \hline
\hline
\end{tabular}
\end{center}
\label{newtab1}
\end{table}

\begin{figure}
\vspace{6cm}
\includegraphics{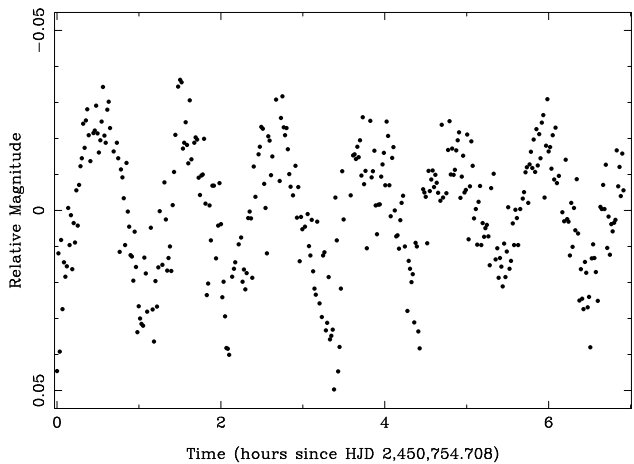}
\caption[]{The white light observations obtained on JD 2,450,754
(1997 November 2 UT) using the 
0.9m telescope. One minute averages of the ten-second integrations
are shown. The zeropoint has been set to the data mean value.}
\label{fig1}
\end{figure}

\begin{figure}
\centering
\centerline{\epsfxsize=8.0cm 
\epsfbox{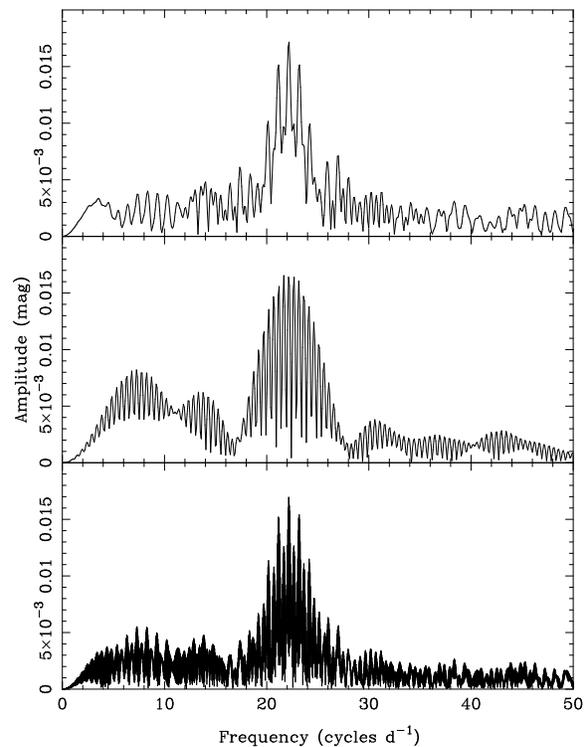}}  
\caption[]{The amplitude spectrum of the all the 0.9m data (top panel);
all the 2.1m data (middle panel); and all the data from both telescopes 
(bottom panel). The observations from each night were individually detrended
by a first order polynomial. In the case of the 2.1m observations, fluxes
from the four channels have been co-added.
}
\label{fig2}
\end{figure}

The photometric runs are cataloged in Table 1;    
approximately 11 and 8 hours of data were obtained on the 0.9m and
2.1m telescopes respectively. The longest continuous light curve is
plotted in Figure \ref{fig1}.  There is a clear modulation with a
period of $\approx 65$ minutes.  Amplitude spectra of various
combinations of photometric runs are shown in Figure \ref{fig2},
plotted over the frequency range of greatest interest. The low
frequency content of the amplitude spectra has been removed to some
extent by detrending the observations linearly. In addition, the fifth
channel observations have been used to correct the data acquired on JD
2450785 for large atmospheric transparency drifts.

\begin{table}
\centering
\begin{center}
\caption[]{A listing of the largest peaks in the amplitude 
spectrum of all the high speed photometric data.}
\begin{tabular}{ccc} \hline
Frequency &    Period & Amplitude \\
 (c/d) &       (days)    & (mmag)    \\
\hline
22.055 & 0.045341 & 14.4 \\
22.086 & 0.045277 & 15.7 \\
22.117 & 0.045215 & 16.6 \\
22.148 & 0.045151 & 17.0 \\
22.178 & 0.045088 & 17.0 \\
22.209 & 0.045027 & 16.6 \\
22.240 & 0.044964 & 15.8 \\
22.271 & 0.044901 & 14.5 \\ \hline
\hline
\end{tabular}
\end{center}
\label{newtab2}
\end{table}

The one cycle per day aliasing pattern is clearly visible in the top
panel of Figure \ref{fig2}, which shows the amplitude spectrum of the
combined 0.9m data (from three successive nights).  The aliasing
pattern is much worse for the combined two 2.1m runs (middle panel),
since these were short. The bottom panel shows the amplitude spectrum
of all the data; although not visible on the scale of Figure
\ref{fig2}, the main peak is composed of an alias pattern of
sub-peaks, spaced 0.031 cycles d$^{-1}$ apart. The frequencies for the
peaks with the largest amplitudes
are listed in Table 2.     
A nonlinear least squares fitting procedure, with a starting guess for
the frequency corresponding to the largest-amplitude alias gave
$f_{\rm max}=22.178\pm 0.00016$ cycles day$^{-1}$ [$P=0.0450897\pm
(3\times 10{-7})$ days] and amplitude $=17.2 \pm 0.3$ mmag.  In spite
of the small formal frequency error, we would hesitate to choose the
correct alias based on the photometry alone.  Some of the reasons for
our misgivings are: the two large peaks in the amplitude spectrum have
almost identical heights; corrections for drifts in the photometry
zeropoints are uncertain; and the 2.1m data used in the procedures
described above was the simple sum of the fluxes in all four filters.
This choice was based on the desire to maximise the available flux,
but implies a distinct mismatch of the wavebands of the two sets of
observations.  It is quite conceivable that the relative heights of
the amplitude spectrum aliases might change if a different combination
of Stiening photometer fluxes is used.  However, we have additional
constraints imposed by spectroscopic observations.  We show below that
the orbital period of KPD 0422+5421 is $P=0.0901795\pm (3\times
10^{-7})$ days, which corresponds to a period of $P=2/f_{\rm max}$.

A linear least squares programme was used to fit the frequency as determined
above to the combined two 2.1m runs; this was done separately for each of
the four Stiening photometer wavebands. The amplitudes 
are given in Table 3.   
Although formal standard errors of the fits are given, these should be
viewed only as very rough guides. In particular, errors in different
waveband datasets are probably highly correlated, so that the evaluation
of e.g. colour phase differences on the basis of the formal errors may be 
misleading.

\begin{table}
\centering
\begin{center}
\caption[]{Amplitudes and phases (with respect to the first
observation) of the variations in the different wavebands. The
contents of this Table were derived by fitting a sinusoid with
$f=22.178$ c/d to the $U$, $B$, $V$ and $R$ data by linear least
squares.  The formal standard errors are given in brackets below each
estimate.}
\begin{tabular}{ccccc} \hline
 & $U$ & $B$ & $V$ & $R$ \\ 
\hline
Amplitude (mmag) & 19.4 & 17.3 & 14.8 & 11.8 \\
(s.e.)    & (0.3)& (0.2)& (0.3)& (0.3)\\ 
          &      &      &      &      \\
Phase    & 1.64 & 1.61 & 1.65 & 1.51 \\
(s.e.)   & (0.02)&(0.01)&(0.02)&(0.03) \\ \hline
\hline
\end{tabular}
\end{center}
\label{tab3}
\end{table}

\section{SPECTROSCOPIC OBSERVATIONS AND ANALYSIS}\label{spectsec}
We obtained 14 spectra of KPD 0422+5421 starting at HJD 2,450,756.8586
(1997 November 4.3583 UT) with the 2.7m telescope at the McDonald
Observatory using the Large Cass Spectrograph, a $600\ell$/mm grating
(blazed at 4200\AA), and the TI1 $800\times 800$ CCD.  The resulting
spectral resolution is $\approx 3.5$~\AA\ (FWHM) with wavelength
coverage of 3525-4940~\AA.  The exposure times ranged from 10 to 20
minutes, and the signal-to-noise ratios in the final reduced spectra
were $\approx 20-40$ in the continuum near H$\beta$.  There were
cirrus clouds present, so the spectra could not be placed on an
absolute flux scale.  Instead, we used nine observations of five
different spectrophotometric flux standards from the nights of
November 1-3, 1997 (which were photometric) to calibrate out the
instrumental response.  Since the slit could not be rotated to follow
the parallactic angle, most of our program objects were observed at
hour angles near zero.  The relative flux scale is reasonably accurate
between H$\beta$ and the Balmer jump---however the relative
calibration at either end of the covered spectral range is suspect in
part because some of the KPD 0422+5421 spectra were obtained at
relatively large airmass (1.5-2.0).

We used the cross-correlation technique of Tonry \& Davis (1979) to
determine the radial velocities of the KPD 0422+5421 spectra.  The
cross-correlation functions (CCFs) were computed over the wavelength
interval 3800-4920~\AA.  We used a synthetic spectrum generated from a
model atmosphere (see below) as the template spectrum (although our
results were almost identical when the first observation was used as
the template).  In each case the CCFs were strong and well-defined.
The velocities corresponding to the centroid of the CCF peaks were
determined by a parabolic fit to the six pixels surrounding the
maximum.

Large radial velocity variations were evident.  We fitted a four-parameter
sinusoid to the 14
radial velocities.  The best-fitting sinusoid had $\chi^2_{\nu}=6.5$, an
indication that the errors on the radial velocities were probably too small. 
We scaled the
errors on the
individual velocities to give $\chi^2_{\nu}=1$ for the
fit.  The resulting spectroscopic elements
are given in Table 4.  
Note that the spectroscopic period of $P_{\rm spect}=0.0907 \pm
0.0020$ days is roughly twice the photometric period derived above.
The spectroscopic period and the velocity semi-amplitude of $K_{\rm
sdB}=237\pm 18$ km s$^{-1}$ imply a mass function of the (unseen)
companion star of
\begin{equation}
f(M)={PK_{\rm sdB}^3\over 2\pi G}=
  {M^3_{\rm comp}\sin^3i\over (M_{\rm comp}+M_{\rm sdB})^2}=0.126\pm 
0.029\,M_{\odot}
\end{equation}
where $M_{\rm comp}$ is the mass of the unseen companion star, 
$M_{\rm sdB}$ is the mass
of the sdB star, and $i$ is the orbital inclination.  The
above equation implies $M_{\rm comp} > 0.126\pm 0.029\,M_{\odot}$.

\begin{table}
\centering
\begin{center}
\caption[]{The orbital parameters for KPD 0422+5421. 
}
\begin{tabular}{cc} \hline
Parameter & Value \\ 
\hline
spectroscopic period (days) & $0.0907\pm 0.0020$ \\
$2\times$photometric   period (days) & $0.0901795\pm (3\times 10^{-7})$ \\
$T_0({\rm spect})$ (HJD 2,450,000+) & $756.9381\pm 0.0010$ \\
$T_0({\rm photo})$ (HJD 2,450,000+) & $785.8199\pm 0.0050$ \\
$K_{\rm sdB}$ velocity (km s$^{-1}$)  &  $237\pm 18$ \\
$\gamma$ velocity (km s$^{-1}$) & $-57\pm 12$ \\
$f(M)$ ($M_{\odot}$)    &   $0.126\pm 0.029$ \\ \hline 
\hline
\end{tabular}
\end{center}
\label{tab4}
\end{table}

\begin{figure}
\centering
\centerline{\epsfxsize=9.0cm 
\epsfbox{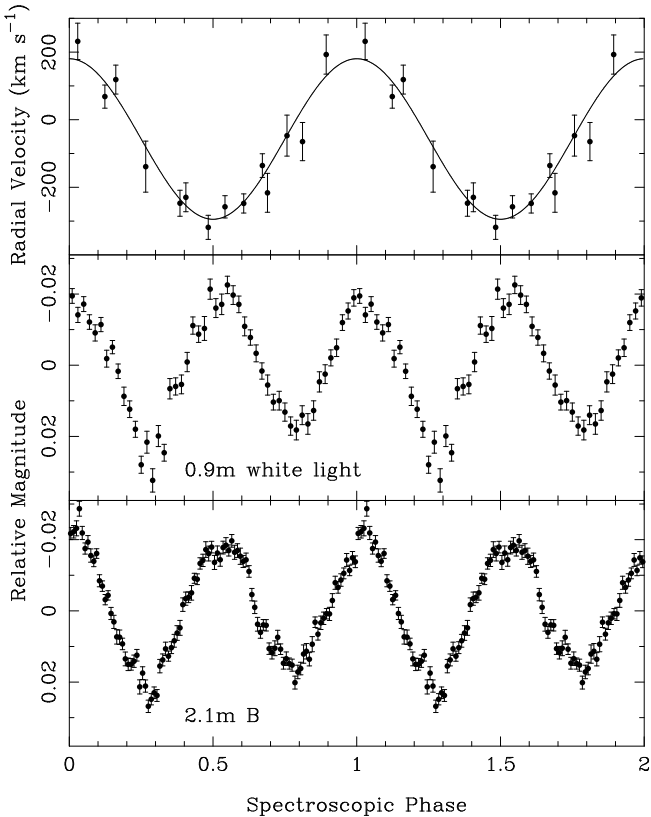}}  
\caption[]{Top: The folded radial velocities of KPD 0422+5421 and the
best-fitting sinusoid.  Middle: The folded light curve from the 0.9m
(white light), binned into 50 phase bins.  Bottom: The folded light
curve from the 2.1m ($B$ band), binned into 100 phase bins.  The
errors shown for the binned light curves are the errors of the mean in
each bin.
}
\label{foldfig}
\end{figure}

The fact that the photometric period is half the spectroscopic period
suggests that the optical modulations we observe are due to the
well-known ``ellipsoidal'' variations.  
The cause of the ellipsoidal
modulations is easy to understand.  The sdB star we observe is
slightly distorted by the unseen companion star.  As the sdB star
moves around in its orbit, its projected area on the sky changes.  The
changes in the projected area give rise to the observed changes in the
total flux.  
We expect to observe the maximum flux at the quadrature
phases when the sdB star is viewed ``side-on.''  Similarly, we will
see minima in the light curve during the conjunction phases when the
sdB star is viewed ``end-on.''  The photometric minimum at the
spectroscopic phase 0.25 (the superior conjunction of the sdB star)
will be slightly deeper due to gravity darkening (see Avni 1978
and the discussion below).  Thus we can use the spectroscopically
determined phase to select the correct alias period from among the
several of the most likely candidates.  We find that only the period
of $2/f_{\rm max}=0.0901795\pm (3\times 10^{-7})$ days correctly
phases the 0.9 and 2.1 meter photometric data with respect to the
radial velocity curve.  The time of the superior conjunction of
the sdB star, as determined from the 2.1m data, is $T_0({\rm photo})
={\rm HJD}\, 2,450,785.8199\pm 0.005$.  The values of 
$T_0({\rm photo})$ and  $T_0({\rm spect})$ differ in phase by
$\Delta\phi=320.270\pm 0.013$, consistent with the expected value of
$\Delta\phi=320.25$.
The folded radial velocity curve and the
folded light curves from the 0.9m (white light) and the 2.1m ($B$) are
shown in Figure \ref{foldfig}.  Models of the ellipsoidal light curve
will be discussed in Section \ref{lcmodels} below.

\begin{figure}
\vspace{6cm}
\includegraphics{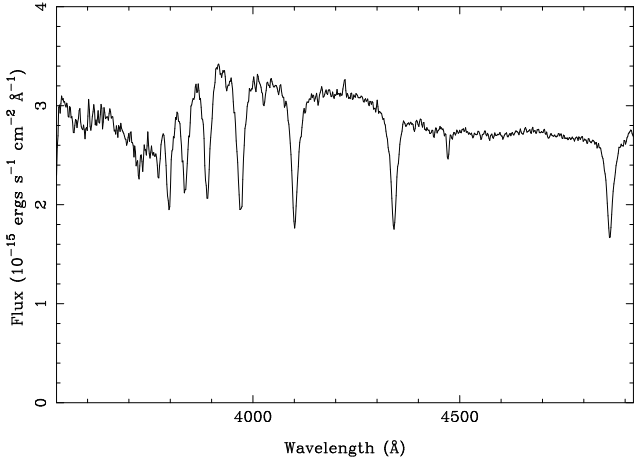}
\caption[]{The rest frame spectrum of KPD 0422+5421, which is the average of
the 14 spectra Doppler corrected to zero velocity.  
The signal-to-noise ratio is about 130 near H$\beta$.
}
\label{plotrest}
\end{figure}

We created a ``restframe'' spectrum of KPD 0422+5421 by Doppler
correcting all 14 spectra to zero velocity and averaging them.  This
restframe spectrum is displayed in Figure \ref{plotrest}.  Two
important points should be noted.  First, the spectrum does not show
any obvious lines from the companion, quite unlike typical
``composite'' spectrum subdwarf stars from the Palomar-Green (PG)
survey (Ferguson, Green, \& Liebert 1984; Green, Schmidt, \& Liebert
1986; Orosz, Wade, \& Harlow 1997).  Second, the He lines are
relatively weak.  We will discuss models of the spectrum in Section
\ref{atmos} below.

\section{CONSTRAINTS ON THE SYSTEM GEOMETRY FROM LIGHT CURVE 
MODELS}\label{lcmodels}
We analyzed the $U$ and $B$ light curves from the observations 
on the 2.1m telescope, by using the
Wilson-Devinney (W-D) code (Wilson \& Devinney 1971; Wilson 1979).  
Only the $U$ and $B$ light curves were used in the modelling process as
these were less noisy than the $V$ and $R$ data. This may be due in part
to the greater difficulty in correcting the observations through the two
red filters for random atmospheric transparancy changes.

We can draw some basic conclusions about the system even before we
begin the modelling.  The mass of the sdB star is probably not very
different from $0.5\,M_{\odot}$ (Saffer et al.\ 1994).  We can compute
the orbital separation as a function of the total system mass from
Kepler's third law:
\begin{equation}
a^3={P^2GM_{\rm total}\over 4\pi^2}.
\label{kepeq}
\end{equation}
We find for KPD 0422+5421 that 
\begin{equation}
{a\over R_{\odot}}=0.846\left({M_{\rm total}\over M_{\odot}}\right)^{1/3}.
\label{areq}
\end{equation}
For mass ratios near 1, the radius of the companion's Roche lobe is
approximately $38\%$ of the orbital separation (Eggleton 1983), which for a
total system mass near $1\,M_{\odot}$, is about $0.35\,R_{\odot}$ for
KPD 0422+5421. If the companion star were on the main sequence, then
it would be later type than about M2 in order to have a radius small
enough to fit inside the Roche lobe (Gray 1992).  Even then, it would
still fill a sizable fraction of its Roche lobe.  Hence we would
expect to see a reflection effect where the irradiation of the M star
by the hot sdB star causes extra light to be observed near the
spectroscopic phase 0.75, as in the case of HW Vir (Wood et al.\
1993).  It is clear from Figure \ref{foldfig} that the light curves
are ellipsoidal---there is no hint of a reflection effect.  This rules
out the possibility that the companion is a cool main-sequence star.
The mass function rules out a brown dwarf of smaller radius.  In
principle, there would not be any sizable reflection effect if the two
stars had nearly the same temperature.  However, if the companion star
had a radius and temperature similar to the sdB star, we would observe
a double-lined spectroscopic binary since $L_{\rm comp}\approx L_{\rm
sdB}$.  The system is single-lined, however.  We conclude that the
companion star is much smaller than the sdB star, and therefore
probably is a white dwarf.

The light curves were fit using ``mode 2'' of the W-D code where
``star 1'' was the sdB star.  Mode 2 is the normal mode used when neither
star is in contact with its Roche lobe.  We assumed circular orbits and
synchronous rotation, both
reasonable assumptions considering the 
proximity of the two stars.  The monochromatic fluxes were computed assuming
blackbodies.  We fixed the polar temperature of the sdB star at
$T_1=26000$~K, and assigned limb darkening
coefficients of $u_1=0.31$ and 0.29 for the $U$ and $B$ bands, respectively
(Wade \& Rucinski 1985).  The sdB star has a radiative envelope, so 
its gravity darkening exponent $g_1$ was set to 1 and its bolometric
albedo $A_1$ for reflective heating and re-radiation was set to 1.  
Since the white dwarf contributes less than $0.1\%$ of the total flux, the
values of $T_2$, $u_2$, $g_2$, and $A_2$ are relatively unimportant.  
Thus we allowed $T_2$ to be a free parameter and set
$u_2=u_1$, $g_2=g_1$, and $A_2=A_1$.  The free parameters in the model
fits are the inclination $i$, the mass ratio $q=M_{\rm comp}/M_{\rm sdB}$,
$T_2$, and the surface potentials $\Omega_1$ and $\Omega_2$.  
The effective wavelengths of the Stiening $U$ and $B$ filters were taken
to be 3460~\AA\ and 4370~\AA, respectively (Wood et al.\ 1993).

\begin{figure}
\centering
\centerline{\epsfxsize=9.0cm 
\epsfbox{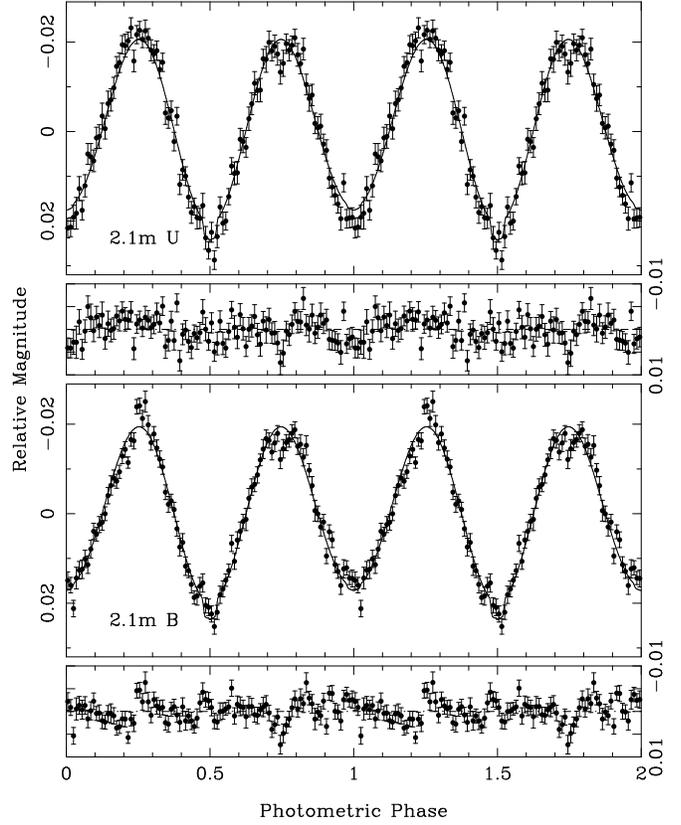}}  
\caption[]{From top to bottom:  The $U$ light curve and the best fit model;
the $U$ residuals in the sense of data minus model; the $B$ light curve and
fit; and the $B$ residuals.  Each point has been plotted twice.
See Table 5 
for the light curve model parameters.
}
\label{lcfits}
\end{figure}

We used a variation of the ``gridls'' program given in Bevington
(1969) to optimize the parameter values.  We defined the optimal light
curve solution to be the one which minimized the $\chi^2$ fit to both
the normalized $U$ and $B$ curves (in magnitude units) simultaneously.
We also used the ``differential corrections'' routine of the W-D code
to check our results.  We are confident we have found the global
$\chi^2$ minimum (rather than a local minimum) 
since we started the optimization routine from a
large number of widely separated regions in parameter space.  We show
in Figure \ref{lcfits} the best fit solution to the $U$ and $B$ light
curves and the residuals.  The overall fit is reasonably good with
$\chi^2_{\rm min} = 364.36$ for 200 data points.  The residuals in the
$U$ and $B$ filters have standard deviations of 0.0029 and 0.0025
magnitudes, respectively.
Table 5 
summarizes the assumed input parameters and the derived
parameters from the model light curve fits.

\begin{figure}
\vspace{6.5cm}
\includegraphics{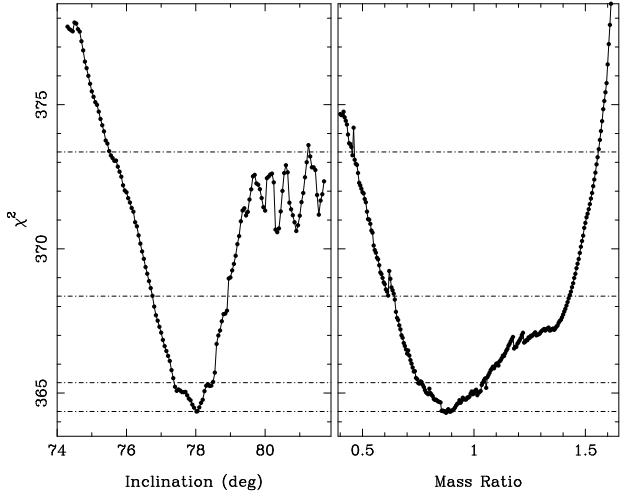}
\caption[]{Left: The $\chi^2$ vs.\ $i$ curve.  The value
of $i$ was fixed at each value along the curve and the other parameters
were adjusted as to give the lowest $\chi^2$.  The dashed-dotted lines
denote $\chi_{\rm min}$, $\chi_{\rm min}+1$, $\chi_{\rm min}+4$, and
$\chi_{\rm min}+9$.  Right: The $\chi^2$ vs.\ $q$ curve.
}
\label{chicut1}
\end{figure}

The statistical errors of the parameters derived from the light curve
models were derived in the following way.  The parameter of interest
(call it $a_i$) was slightly altered from its optimal value
($a^1_i=a^0_i+\delta$) and held fixed while the other free parameters
were adjusted to give the minimum $\chi^2$.  The process was repeated
($a^2_i=a^1_i+\delta$, etc.) until $\chi^2$ differed substantially
from $\chi^2_{\rm min}$.  Figure \ref{chicut1} shows the $\chi^2$ vs.\
$i$ and the $\chi^2$ vs.\ $q$ curves.  A change in the inclination $i$
of $\approx 0.5^{\circ}$ is required to force $\chi^2$ to change by 1.
We adopt $\sigma_i=0.50^{\circ}$, although we note that this is a
rough approximation since the curve is not parabolic and $\chi^2_{\rm
min}$ is much larger than the number of data points fit.  Based on the
$\chi^2$ vs.\ $q$ curve, we adopt $\sigma_q=0.15$ for the sake of
discussion (and with the same caveats discussed for $\sigma_i$).  The
quoted errors on the other parameters given in Table 5 
were derived assuming Gaussian errors.  Eggleton's (1983) formula was
used to compute the effective radii of the Roche lobes
\begin{eqnarray}
{R_{Rl}({\rm sdB})\over a}
         &=&{0.49q^{-2/3}\over 0.6q^{-2/3}+\ln (1+q^{-1/3})}  \nonumber \\
{R_{Rl}({\rm comp})\over a}
          &=&{0.49q^{2/3}\over 0.6q^{2/3}+\ln (1+q^{1/3})}.
\label{eggeq}
\end{eqnarray}
The W-D code computes the radii of the components using the values
of $\Omega_1$, $\Omega_2$, and the orbital separation $a$.
The sdB star is well within its
Roche lobe.

We tried model fits with the temperature of the sdB star set to 25,000~K
and to 27,000~K.  The best-fit values of $i$ and $q$ were nearly identical
to those values given in Table 5.  In that same vein,
we also tried model fits where the limb darkening coefficients 
were set to 0.25 and 0.35 for both filters.  Again, the best-fit values
of $i$ and $q$ did not change significantly.  Finally, we found that 
the best-fit values of $i$ and $q$ did not change significantly when
the effective wavelengths of the $U$ and $B$ filters were changed by
100~\AA.  We conclude that our results presented in Table 5 are 
insensitive to the exact values chosen for the sdB parameters.

\begin{table}
\centering
\begin{center}
\caption[]{Fitted parameters for KPD 0422+5421.}
\begin{tabular}{cc} \hline
Parameter & Value \\ 
\hline
$i$ (deg)  &   $78.02\pm 0.50$  \\
$q$        &   $0.87\pm 0.15$   \\
$T_2$ (K)  &   $\approx 4040$    \\
$\Omega_1$ &   $4.8642\pm 0.24$           \\
$\Omega_2$ &   $\approx 57.6086$          \\
           &                    \\
$M_{\rm sdB}$ ($M_{\odot}$) &    $0.72 \pm 0.26$  \\
$M_{\rm comp}$ ($M_{\odot}$) &    $0.62 \pm 0.18$  \\
$R_{\rm sdB}$ ($R_{\odot}$) &    $0.248 \pm 0.008$  \\
$R_{\rm comp}$ ($R_{\odot}$) &    $\approx0.01$  \\
$\log g_{\rm sdB}$ (cgs)    &   $5.50\pm 0.21$  \\
$\log g_{\rm comp}$ (cgs)    &   $\approx 7.9$  \\
                            &                   \\
$R_{Rl}$(sdB) ($R_{\odot}$) &    $0.36\pm 0.05$ \\
$R_{Rl}$(comp) ($R_{\odot}$) &    $0.35\pm 0.03$ \\
expected $V_{\rm rot}\sin i$ (km s$^{-1}$) & $137\pm 4$ \\
$a$           ($R_{\odot}$) &    $0.93\pm 0.11$  \\  \hline
\hline
\end{tabular}
\end{center}
\label{tabfit}
\end{table}

We emphasize that we have only estimated the statistical errors and not
any systematic errors.  There are no doubt systematic errors present, given
the small light curve amplitudes and the nature of the
observations (i.e.\ high speed  photometry).  However,
the white light curve from the 0.9m and the $B$ light curve from the
2.1m are very similar in amplitude, and their relative phases are as
expected.  
There is also nothing unusual about the derived component masses.
The mass of the sdB star ($0.72 \pm 0.26\,M_{\odot}$)
is consistent with the ``canonical'' extended horizontal branch
mass of $0.5\,M_{\odot}$
(Caloi 1989; Dorman, Rood, \& O'Connell 1993; Saffer et al.\ 1994).  
Webbink (1990, see also Warner 1995)
finds a mean white dwarf mass of $0.74\pm 0.04\,M_{\odot}$ in 84
cataclysmic variable binaries, and the mean mass of
white dwarfs in the field is near $0.56\,M_{\odot}$
(Bergeron, Saffer, \& Liebert 1992). 
The mass of the white dwarf in KPD 0422+5421
($0.62 \pm 0.18\,M_{\odot}$) is consistent with either of these. 
These considerations suggest that the systematic errors are
reasonably small and that ou derived parameters are trustworthy.

A close examination of the model light curve near the photometric phase
0.5 (when the sdB star is {\em behind} the companion) shows that the
white dwarf companion passes in front of the sdB star, causing a dip of
$\approx 0.005$ magnitudes which lasts $\approx 0.05$ in phase
(6.5 minutes).  There is also a very slight distortion at phase
1.0 where the nearly invisible white dwarf is totally eclipsed by the
sdB star.  Interestingly enough, the white-light curve from the 0.9m
shows a possible indication of a small dip near the spectroscopic phase
0.25 (i.e.\ the photometric phase 0.5). 
Caution suggests that the observation of a transit be considered only
tentative.  Since the depth of the depression caused by the transit
is only $\approx 0.005$ magnitudes, its exact shape is quite
sensitive to slight drifts in the transparency, etc.  Extended observations
with a CCD on a large telescope (where one can use reasonably short exposures)
should be done to confirm the existence of the transit.  The use of
a CCD enables one to do precise differential photometry over many orbital
cycles.

\section{SUBDWARF ATMOSPHERIC PARAMETERS}\label{atmos}
The effective temperature $T_{\rm eff}$ and surface gravity $\log g$
of the subdwarf were determined using a grid of
synthetic spectra.  LTE hydrogen and helium line-blanketed model
atmospheres were constructed using version 178 of the atmosphere code
{\sc TLUSTY} (Hubeny 1988), for solar abundances and for reduced helium
abundance, [He/H] = $-0.5, -1.0$, and $-1.5$ (where square brackets
denote the logarithmic number ratio relative to solar).  Detailed
spectra were synthesized on the interval 3780 -- 4910 \AA\ using
version 41 of {\sc SYNSPEC} (Hubeny, Lanz \& Jeffery 1994); H and He I lines
were computed using line broadening tables, while lines from other
elements were computed using Voigt profiles and data from the Kurucz
atomic line list.  A microturbulent velocity parameter of 4 km~sec$^{-1}$ was
assumed. An approximate NLTE treatment of line opacity based on
second-order escape probability was used in the spectrum synthesis
(see Hubeny, Harmanec \& Stefl 1986).  The spectra were then convolved
and sampled to match the dispersion (1.77 \AA\ pixel$^{-1}$) and
instrumental resolution (3.5 \AA).  On the assumption that the
subdwarf's rotation is tidally synchronized to the binary orbital
period, we included a rotational broadening of 
$V_{\rm rot}\sin i = 150$~km~sec$^{-1}$, which is
less than the FWHM of the instrumental profile.  At each value
of [He/H], models and spectra were computed at spacings of 1000~K for
$T_{\rm eff} = 23000$ -- $28000$~K, and spacings of 0.2 for $\log g =
5.0$ -- $5.8$.  For [He/H] $< 0$, the grid resolution was increased in the
neighborhood of the best fit, to 500~K for $T_{\rm eff} = 24000$ --
27000~K and 0.1 for $\log g = 5.2$ -- 5.6.

\begin{figure}
\vspace{6.5cm}
\includegraphics{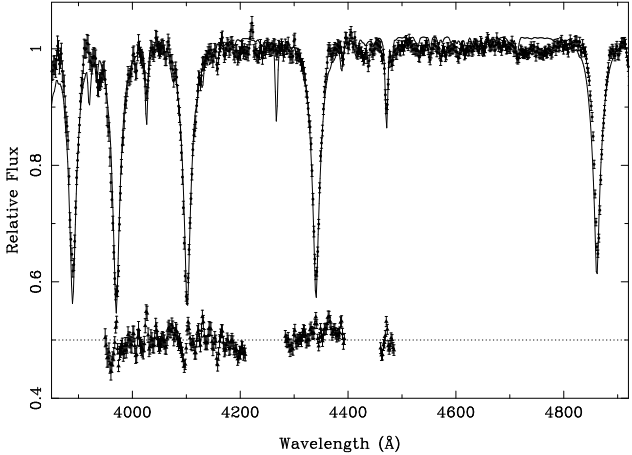}
\caption[]{The restframe spectrum (filled circles)
is displayed with the best-fitting model (solid line) and the
residuals in the sense of data minus model (filled triangles), offset by 0.5
units upward.  The CII line at 4266.7~\AA\ was not included in the fitting
region.
}
\label{fitplot}
\end{figure}

The models and the restframe spectrum were normalized to their
continua, and the $\chi^2$ and rms residuals were computed using most
of the spectrum between 3900~\AA\ and 4900~\AA.  The H$\beta$ profile
and nearby continuum were not well fit by the models, in part due to
the relatively poor calibration near the ends.  The CII line at
4266.7~\AA\ is much weaker in the data than in the model spectra.
After some experimentation, final fitting regions were chosen to be
$3949.0\le\lambda\le 4210.0$, $4282.0\le\lambda\le 4394.0$, and
$4460.0\le\lambda\le 4486.0$.  This excludes the CII line, some bad
pixels near 4222~\AA, and the region near 4430~\AA\ which has a weak
interstellar absorption band.  The best-fitting model (Figure
\ref{fitplot}) has $T_{\rm eff}=25,000$~K, $\log g=5.4$, and [He/H] =
$-1.0$.  
The rms residual of the fit is 0.018 (unit = normalized
continuum).  For comparison, the value of $\log g$ derived from the
photometrically determined mass and radius of the sdB star is
$\log g=5.50\pm 0.21$ (Table 5).
Keeping $T_{\rm eff}$, $\log g$, and [He/H] at their best
values, we find that an approximate match of the models to the
strength of the CII line at 4266.7~\AA\ requires a reduction in the
carbon abundance by a factor of ten or more with respect to solar.
Our resolution and signal-to-noise ratio are insufficient to make
further deductions about abundances of metals.

The estimated sizes of the $1\sigma$ statistical errors are
$\sigma_{T}\approx 1500$~K and $\sigma_{\log g}\approx 0.1$.  There is
a positive correlation between the estimated parameters $T_{\rm eff}$
and $\log g$, in the sense that forcing a displacement of 0.1 dex in
$\log g$ requires a displacement of $\sim 700$~K to minimize $\chi^2$.
Forcing [He/H] to a value away from the best fitting $-1.0$
pushes $T_{\rm eff}$ and $\log g$ along this correlation.  Our values
of $T_{\rm eff}=25,000\pm 1500$~K and $\log g=5.4\pm 0.1$ are
consistent with those values communicated by Saffer ($T_{\rm
eff}=26,050$, $\log g=5.51$, [He/H] unknown).  If we had used {\sc SYNSPEC}
to generate synthetic spectra using line-blanketed models from Kurucz
(1994 CD-ROM distribution), we estimate that our inferred $T_{\rm
eff}$ would be $\sim 1500$~K lower with no change in $\log g$, based
on comparisons made at [He/H]= 0.  We stress that we have not
calibrated our parameter fitting procedure directly against stars
analysed by other workers.

\begin{figure}
\vspace{6.5cm}
\includegraphics{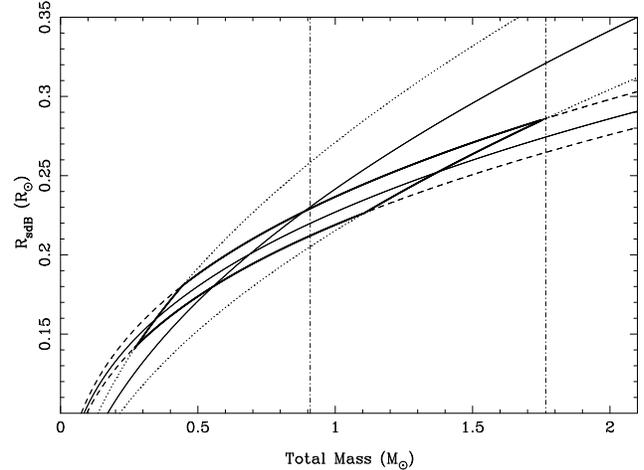}
\caption[]{The relations given in Equations \protect\ref{gradeq}
and \protect\ref{kepradeq}
are plotted as the solid lines.  The dotted lines
show Equation \protect\ref{gradeq} using the $1\sigma$ errors
on $\log g$.  The dashed lines show
Equation \protect\ref{kepradeq} using the $1\sigma$ errors
on $q$  The region where the two $1\sigma$ regions overlap
is outlined with the thick line.  Finally, the $1\sigma$
range of $M_{\rm total}$ allowed by the mass function is shown with the
vertical dashed-dotted lines.
}
\label{radiusplot}
\end{figure}

The spectroscopic value of $\log g$ can be used to make an independent
estimate of the radius of the sdB star.  As a function of the total mass
the radius of the sdB star is
\begin{equation}
{R_{\rm sdB}\over R_{\odot}}=\left[
      \left({M_{\rm total}\over M_{\odot}}\right)
      \left({g_{\odot}\over g}\right)
      \left({1\over 1+q}\right)\right]^{1/2}
\label{gradeq}
\end{equation}
where  $q \equiv M_{\rm comp}/M_{\rm sdB}$ and
$M_{\rm total} \equiv M_{\rm sdB}+M_{\rm comp}$.  
On the other hand, the radius of the sdB star is given by the light
curve solution as a fraction ${\cal F}$ of the effective Roche lobe
radius $R_{\rm Rl}(q)$.  Thus using Kepler's third law,
\begin{equation}
R_{\rm sdB}={\cal F}R_{\rm Rl}(q)
  \left[{P^2GM_{\rm total}\over 4\pi^2}\right]^{1/3}
\label{kepradeq}
\end{equation}
where $R_{\rm Rl}(q)$ is given in Equation \ref{eggeq}.  In
Figure \ref{radiusplot} we plot equation \ref{gradeq} using $\log g=5.4$ and
Equation \ref{kepradeq} using $q=0.87$.  The two curves cross at
$M_{\rm total}\approx 0.7\,M_{\odot}$.  We also show the $1\sigma$
limits on each curve ($\sigma_{\log g}=0.1$ and $\sigma_q=0.15$).  The
region where the two error bands overlap is outlined with the thick lines.
An extreme range of $0.25\le M_{\rm total} \le 1.75\,M_{\odot}$ is
allowed, compared with the $1\sigma$ range of $0.91\le M_{\rm total} \le
1.77\,M_{\odot}$ allowed by the mass function and the photometrically
determined values of $q$ and $i$.  We conclude that the radius derived
from our spectroscopic value of $\log g$ is basically compatible with
the radius derived from the photometric determination of $q$ and $i$,
while favoring lower values of $M_{total}$.

A measurement of the projected rotational velocity of the sdB star
$V_{\rm rot}\sin i$ would provide another measurement of its radius:
\begin{equation}
R_{\rm sdB}=V_{\rm rot}\sin i\left({P\over 2\pi\sin i}\right),
\label{vroteq}
\end{equation}
provided the sdB star is synchronously rotating.  Our present spectral
resolution is not sufficient to measure $V_{\rm rot}\sin i$ for KPD
0422+5421 directly. Such a measurement 
would provide another useful constraint on the total system mass
through the relations given in Equations
\ref{gradeq} and \ref{kepradeq}.

\section{The 
REDDENING FROM ABSOLUTE PHOTOMETRY
and the distance}\label{red}
Absolute photometry of KPD 0422+5421 in the Johnson $UBV$ and
Str{\"o}mgren systems was obtained by Downes (1986) and Wesemael et
al. (1992) respectively. The results are: $V=14.66$, $B-V=+0.20$,
$U-B=-0.65$ (no errors are quoted); and $y=14.682\pm 0.018$,
$b-y=0.141\pm 0.011$, $u-b=0.477\pm 0.001$, $m_1=0.063\pm 0.019$.
Since the photometric variations are quite small and in phase at
different wavelengths, it may be assumed that the accuracy of the
absolute photometry is largely unaffected by the cyclical variations.

Comparison of the $U-B$ and $B-V$ indices with those of a large
collection of hot subdwarfs given in Fig.\ 2 of Kilkenny et
al. (1997b) makes it clear that KPD0422+5421 is anomalously red. In
fact, its colours strongly resemble those of hot, subluminous stars
with composite spectra, as shown in Fig.\ 3 of Kilkenny et
al. (1997b). However, our spectra do not exhibit any signs of anything
but the sdB star.  Furthermore, our photometric solution requires the
companion to be degenerate.  The obvious conclusion is that
KPD0422+5421 is substantially reddened.

Further support for this deduction comes from the Str{\"o}mgren colour
indices quoted above. Wesemael et al. (1992) plot reddening lines in a
$(u-b, b-y)$ diagram, for a range of assumed sdB temperatures (their
Fig.\ 9); KPD0422+5421 inhabits an area in the diagram with
$E(b-y)>0.2$. A crude estimate gives $E(b-y)\approx 0.25$, which is
not surprising given that the star is in the galactic plane.  By
comparing the observed Str{\"o}mgren photometric indices to stellar
atmosphere model predictions, Villeneuve et al. (1995) derive $E(b-y)
\approx 0.26$.  The value of $E(b-y)=0.25$ implies $E(B-V)=0.36$
(Crawford, Glasky, \& Perry 1970).

To compute the distance to the source, we adopt a temperature of $T_{\rm
eff}=25,000\pm 1500$~K, a radius of $R_{\rm sdB}=0.249\pm
0.008\,R_{\odot}$, a colour excess of $E(B-V)=0.36\pm 0.05$, a
bolometric correction of $BC=2.5\pm 0.1$ (Gray 1992), and an apparent
$V$ magnitude of $V=14.66$ (Downes 1986).  We find $L_{\rm
bol}=21.8\pm 5.4\,L_{\odot}$, $M_V=3.86\pm 0.29$, and $d=850\pm 130$
pc.  For comparison, Villeneuve, Wesemael, \& Fontaine (1995) find
$581 \le d \le 1120$~pc based on temperatures derived from
Str{\"o}mgren photometry.

\section{DISCUSSION OF SIMILAR SYSTEMS AND EVOLUTIONARY 
CONSIDERATIONS}\label{evol}

Three issues relating to the binary evolution arise from our
study of KPD 0422+5421.  

First, we note that the sdB star, if interpreted as a member of the
extreme horizontal branch (EHB), will soon evolve to become the second
white dwarf  in KPD 0422+5421.  
Thus KPD 0422+5421 is a predecessor to the class
of double degenerate binaries.  Presumably it has already passed
through a common envelope phase.  A possible predecessor in turn to
KPD 0422+5421
is a system like HD 185510 (V1379 Aql), which is a 21-day binary
containing an evolved compact (sdB) star and a K0 giant (Fekel et al.\
1993; Jeffery \& Simon 1997).  The mass ratio of HD 185510 is large,
$M_{\rm K}/M_{\rm sdB} 
\approx 7.5$ (Fekel et al.\ 1993), and one may expect that
a common envelope stage lies ahead.
At that time, HD 185510 may resemble KPD 0422+5421, with its present sdB star
in the r\^ole of KPD's white dwarf.

Second, what is the present total binary mass?  The combined
spectroscopic and photometric orbital solution suggests a value
tantalizingly close to the Chandrasekhar limit, $M_{\rm Ch} \approx
1.4\,M_{\odot}$, with all that implies for the possibility that 
KPD 0422+5421 may
become a Type Ia supernova, when the components merge.  On the other
hand, if the sdB star is truly a core-He burning EHB star, its mass is
expected to be close to $0.5\,M_{\odot}$, 
and with $q$ derived from the light
curve solution constraining the mass of the unseen white dwarf, the
total mass may be considerably less than $M_{\rm Ch}$.  Improved light
curves and radial velocity data will help to address this question.
In the meantime, it is well to bear in mind the discussion in Jeffery
\& Simon (1997) concerning whether the canonical $0.5\,M_\odot$ for EHB
stars is really empirically well-determined.

Third, short-period binary systems of the same type as KPD 0422+5421, 
namely sdB
+ white dwarf, 
are rare, and KPD 0422+5421 arguably possesses the shortest period of them
all. [One double degenerate system has a shorter period:
WD 0957-666, 
$P = 0.060993$ days, (Moran, Marsh,
\& Bragaglia 1997). The total mass is estimated to be only $0.69\,M_{\odot}$, 
however.]
The catalog of Ritter \& Kolb (1998) lists three possible
sdB+``white dwarf'' pairs (i.e.\ sdB stars known to be in close
binaries with undetected companions), and the sample of Saffer, Livio,
\& Yungelson (1998) contains seven sdB binaries with undetected
companions (including two from Ritter \& Kolb's catalog), of which
five have known orbital periods.  The six objects with known orbital periods 
(in order of increasing period)
are PG 1432+159 ($P=5.39$
hours), PG 2345+318 ($P=5.78$ hours), Feige 36 ($P=8.5$ hours), 
PG 0101+039 ($P=13.7$ hours), HZ 22
(=UX CVn, $P=13.77$ hours),  and Ton 245 ($P=2.5$
days).  A related sdO + white dwarf(?) system is HD 49798 with
$P=1.55$ days (Thackeray 1970).
The best known member of the sdB + white dwarf class is HZ 22 (Humason \&
Zwicky 1947; Young, Nelson, \& Mielbrecht 1972; Greenstein 1973).
Greenstein (1973) and Young \& Wentworth (1982) argue that the unseen
companion of HZ 22 is a white dwarf.  However, it is also possible
that HZ 22 contains a low mass main sequence star companion, so the
nature of HZ 22 is still an open question.  Another recent sdB+``white
dwarf'' candidate is V46 in the globular cluster M4 (Kaluzny, Thompson
\& Krzeminsky 1997), which is an sdB star that shows an apparently
sinusoidal light curve with a period of 1.045 hours.  This period is
much longer than those seen in the pulsating sdB stars (EC14026 stars,
Kilkenny et al. 1997a).  However, the nature of the companion star is
not known and it is not clear if the photometric period is the orbital
period.  Since it has the shortest confirmed orbital period,    
KPD 0422+5421 
will evolve more quickly {\it via\/} gravitational wave radiation
of angular momentum than the other sdB + white dwarf systems.
Ritter (1986) gives the time
required for a detached binary to reach the semi-detached state:
\begin{equation}
t_{\rm sd} =(4.73\times 10^{10}\,{\rm yr}){M_{\rm total}^{1/3}\over
  M_{\rm comp}M_{\rm sdB}}P^{8/3}
  \left[1-\left({P_{\rm sd}\over P}\right)^{8/3}\right]
\end{equation}
where the masses are in solar units, the periods are in days, and where
\begin{equation}
{P_{\rm sd}\over P}=\left({R_{\rm sdB}\over R_{Rl}({\rm sdB})}\right)^{3/2},
\end{equation}
which assumes that the radius of the sdB star stays fixed.  Using
$P_{\rm sd}/P=0.57\pm 0.13$, we find that $\log t_{\rm sd}=8.17\pm
0.15$ ($t_{\rm sd}\approx 1.5\times 10^8$ years), which is comparable
to the core He burning lifetime of an sdB star ($\approx1.5\times
10^8$ years, Dorman, Rood, \& O'Connell 1993).
Thus, possibly before the sdB star has evolved to a white dwarf, there
will be a further episode of mass exchange. What KPD 0422+5421 will then
look like, and whether it can be identified with any presently known class of
objects, takes the subject well beyond the expertise or inclination
of the present authors to speculate.



\section{Summary}
The sdB star KPD 0422+5421 was found to be a short period detached
binary with a period of $P=0.0901795\pm (3\times 10^{-7})$ days.  We
argue that the companion star is a white dwarf since the companion
star is not seen in the spectra, and the synthetic light curve model
of the $U$ and $B$ light curves requires its radius to be on the order
of $\approx 0.01\,R_{\odot}$.  
We
derive component masses of $M_{\rm
sdB}=0.72\pm 0.26\,M_{\odot}$ and $M_{\rm comp}= 0.62\pm
0.18\,M_{\odot}$. 
KPD 0422+5421 
is
one of a small number of known sdB + white dwarf binary systems and
represents a poorly
observed and short-lived stage of binary star evolution.

\section*{Acknowledgments}
The authors are grateful to Dr.\ Rex Saffer for supplying his gravity
and temperature estimate of KPD 0422+5421 before publication.
We also thank Peter Eggleton for pointing out to
us the intriguing and possibly related system HD 185510.  CK
thanks Prof.\ Ed Nather for his hospitality at the University of Texas
(Austin); the director of the McDonald Observatory for a generous
allotment of telescope time; Rob Robinson and Rae Stiening for the use
of the Stiening photometer; and the South African Foundation for
Research Development for partially funding some of this work.  JO and
RW acknowledge partial support from NASA grant NAG 5-3459 to the
Pennsylvania State University.  We are grateful to Ivan Hubeny for
instruction in the use of his stellar atmosphere codes {\sc TLUSTY},
{\sc SYNSPEC}, and {\sc ROTINS}.  This research has made use of the Simbad
database, operated at CDS, Strasbourg, France.

\bsp

\label{lastpage}

\end{document}